\documentclass[aip, amsmath,amssymb, reprint,]{revtex4-1}

\usepackage[utf8]{inputenc}
\usepackage{makecell}
\usepackage{array}
\usepackage{dcolumn}
\usepackage{physics}
\usepackage{lmodern}
\usepackage{graphicx}
\usepackage{amsmath}
\usepackage{amssymb}
\usepackage{float}
\usepackage{color}
\usepackage[dvipsnames]{xcolor}
\usepackage{xspace}
\usepackage[version=4]{mhchem}
\usepackage[T1]{fontenc}
\usepackage[normalem]{ulem}

\graphicspath{{figures/}}

\newcommand{\etal}{\textit{et al.}\xspace}

\newcommand{\abinitio}{\textit{ab initio}\xspace}

\newcommand{\ANG}{\text{\AA}\xspace}

\usepackage{silence}
\begin{document}


\title{Equation of state of solid parahydrogen using \abinitio two-body and three-body interaction potentials}

\author{Alexander Ibrahim}
\affiliation{Department of Physics and Astronomy, University of Waterloo, 200 University Avenue West, Waterloo, Ontario N2L 3G1, Canada}
\affiliation{Department of Chemistry, University of Waterloo, 200 University Avenue West, Waterloo, Ontario N2L 3G1, Canada}

\author{Pierre-Nicholas Roy}
\email{pnroy@uwaterloo.ca}
\affiliation{Department of Chemistry, University of Waterloo, 200 University Avenue West, Waterloo, Ontario N2L 3G1, Canada}


\begin{abstract}

We present the equation of state (EOS) of solid parahydrogen
between $ 0.024 \, \ANG^{-3} $ and $ 0.1 \, \ANG^{-3} $
at $ T = 4.2 $ K,
calculated using path-integral Monte Carlo simulations,
with \abinitio two-body and three-body interaction potentials.
We correct for finite size simulation errors using potential tail corrections.
Trotter factorization errors are accounted for,
either via extrapolation,
or by using a suitably small imaginary time step.
We incorporate the three-body interaction using two methods;
the full inclusion method,
where pair and three-body interactions are used
in both Monte Carlo sampling and in the energy estimators,
and the perturbative method,
where three-body interactions are omitted from sampling
but are still present in energy estimations.
Both treatments of the three-body interaction
return very similar total energies and pressures.
The presence of three-body interactions has only minor effects
on the structural properties of the solid.
Whereas the pair interaction,
on its own,
significantly overestimates the pressure of solid parahydrogen,
the additional presence of the three-body interaction
causes a severe underestimation of the pressure.
Our findings suggest that
accurate simulations of solid parahydrogen
require four-body and possibly higher-order many-body interactions.
It may also be the case that
static interaction potentials are entirely unsuitable
for simulations of solid parahydrogen at high densities.

\end{abstract}

\maketitle


\section{Introduction} \label{sec:intro}
Solid parahydrogen is the simplest quantum molecular solid that exists in nature.
It is the target of a sizeable body of active research.\cite{ph2solidexp:21bhan, ph2solidexp:21mutu, ph2solidexp:22shel, ph2solidexp:20stro, ph2solidtheo:21boni, ph2solidtheo:17duss}
As a quantum solid,
solid parahydrogen possesses many interesting and exotic physical properties.
It has a triple point of
$ (T, P) = (13.8 \, \mathrm{K}, 7.042 \, \mathrm{kPa}) $.\cite{moleculeh2:09leac}
Upon freezing,
solid parahydrogen exists in its stable equilibrium form
as a hexagonal close-packed (\textit{hcp}) lattice.
Due to its low mass,
each parahydrogen molecule has a very large zero-point energy,
and is very delocalized about its nominal lattice site.
At the experimental equilibrium density of $ \rho_0 = 0.0260 \, \ANG^{-3} $,
the Lindemann ratio of solid parahydrogen is about $0.2$.\cite{ph2solidexp:12fern, ph2solidtheo:17duss}
These zero-point motions expand the lattice to the extent that,
at the equilibrium density,
the parahydrogen molecules are pushed to a mean distance beyond the pair potential's minimum.
As a result, solid parahydrogen is highly compressible.\cite{ph2solidexp:80silv}

As a molecular solid,
a key property of solid parahydrogen
is that the individual parahydrogen molecules within
retain most of their free molecule properties.
The interactions of an individual parahydrogen molecule
with its surroundings in the solid only slightly perturb its behaviour
from that of a free molecule.
For example, except at very high pressures,
the free molecule rotational and vibrational quantum numbers $ j $ and $ \nu $
remain good quantum numbers in the solid.
We can readily model solid parahydrogen
using interactions between collections of free parahydrogen molecules,
such as pairs and triplets.

Because of its simplicity and relative ease of modelling
compared to more complicated systems,
solid parahydrogen is a convenient benchmark against which
we can test our current theoretical knowledge of quantum many-body physics.
Constructing the equation of state (EOS)
is an excellent way to test the accuracy
of a potential energy surface (PES) for parahydrogen under different density and state conditions.

Most simulations of bulk condensed parahydrogen are done
using static point-wise pair interaction potentials.
Such potentials have many advantages.
They are computationally fast and easy to implement;
a dense lookup table and linear interpolation are often enough.
The Silvera-Goldman (SG) and Buck potentials
are two of the most popular two-body potentials.\cite{h2pes:78silv, h2pes:83buck}
Both provide very good predictions at low and moderate densities,
but overestimate pressures at higher densities.\cite{h2pes:13omiy}
The Faruk-Schmidt-Hinde (FSH) potential is a more recent
\abinitio first principles pair potential,\cite{ph2cluster:14faru, ph2cluster:15schm}
and due to its construction,
it contains no implicit many-body contributions.
Path-integral Monte Carlo (PIMC) simulations show that,
even at relatively low densities (below $ 0.04 \, \ANG^{-3} $)
the FSH potential overestimates the pressure of solid parahydrogen
even more drastically than do the SG and Buck potentials.\cite{pathinteg:19ibra}

Each of the aforementioned SG, Buck, and FSH potentials
has a core wall that is too rigid,
causing them to overestimate the pressure of solid parahydrogen at high densities.
In 2013,
Omiyinka and Boninsegni\cite{h2pes:13omiy}
modified the Moraldi potential\cite{h2pes:12mora} to create a new pair potential,
whose highlighting feature is its drastically softened core wall
when compared to the SG potential.
Simulations using this pair potential
reproduce the experimental pressure-density curve for parahydrogen remarkably well
between the densities of $ 0.076 \, \ANG^{-3} $ and $ 0.273 \, \ANG^{-3} $.

Several computational studies of condensed systems of parahydrogen
emphasize the importance of three-body interactions at high densities.\cite{threebody:08hind, threebody:10manz, h2pes:13omiy, pathinteg:19ibra}
These three-body interactions are attractive at short intermolecular separations,
and their inclusion 
effectively softens the core wall of the pair potential.
The Axilrod-Teller-Muto (ATM) potential is a commonly-referenced potential
for triple-dipole interactions.\cite{threebody:43axil}
Unfortunately,
despite having the correct long-range behaviour,
it fails to reproduce the desired short-range behaviour even qualitatively.
The ATM potential predicts a net repulsive interaction at high densities,
the opposite of what is required.

This research group recently published
a static \abinitio three-body interaction potential energy surface (PES)
for parahydrogen.\cite{threebody:22ibra}
This new PES has the expected net attractive interaction at short ranges,
as well as the expected long-range behaviour of the ATM potential.

We can also look at studies of a closely related system,
solid helium-4,
for additional insights on solid parahydrogen.
Solid helium-4
is a quantum molecular solid above $ 25 $ bar,
its atoms have large zero-point motions
and interact weakly through dispersion interactions,
and, like solid parahydrogen,
it is commonly simulated using quantum Monte Carlo techniques.
In 2017, Barnes and Hinde performed variational Monte Carlo (VMC) and path-integral Ground State (PIGS) simulations
of solid helium-4.\cite{pathinteg:17barn_1}
They used the Aziz-87 pair potential,\cite{he4pes:87aziz}
and the three-body potential of Cencek \etal\cite{he4pes:09cenc}
Simulations using both the pair and three-body potentials
reproduce the experimental pressure-volume data
down to about $ 4.0 \, \mathrm{cm}^{3} / \mathrm{mol} $,
whereas those with only the pair potential
were successful only down to about $ 8.0 \, \mathrm{cm}^{3} / \mathrm{mol} $.
Given the similarities between solid helium-4 and solid parahydrogen,
the results of Barnes and Hinde are an indication
that three-body interactions may greatly affect
the EOS of solid parahydrogen.

In this work, we compute the EOS of solid parahydrogen at $ T = 4.2 $ K
using PIMC simulations.
The system is constructed as an \textit{hcp} lattice
between the densities of $ 0.024 \, \AA^{-3} $ and $ 0.1 \, \AA^{-3}$.
We use a potential tail calculation,
for both the pair and three-body potentials,
to correct the finite system size error.
We also explore how these tail corrections vary as a function of density.
Depending on the requirements,
as described in Sec.~\ref{sec:computation},
we correct the Trotter factorization error
either via extrapolation,
or by choosing a suitably small imaginary time step.

Following the work by Barnes and Hinde\cite{pathinteg:17barn_1}
we explore two different ways of including the effects of the three-body interaction
into our simulations.
The first is the full inclusion method,
in which both the pair and three-body potentials
are used for the Monte Carlo sampling as well as in the energy estimators.
The second is the perturbative method,
in which only the pair potential is used for the Monte Carlo sampling,
but both potentials are used in the energy estimators.
The key draw of the perturbative method
is that it drastically reduces the computational costs of sampling.
The perturbative case also gives us,
with no additional effort,
the EOS for which only the pair potential is used in the estimator.
We also look at how
each of these two methods of incorporating the three-body interaction
affects the structure of solid parahydrogen.
For brevity,
in this paper we will refer to ``simulations performed with the perturbative treatment of the three-body potential'' as ``perturbative treatment simulations'',
and ``simulations performed with the full inclusion treatment of the three-body potential'' as ``full inclusion treatment simulations''.

The remainder of this paper is structured as follows.
In Sec.~\ref{sec:model} we present the model of the system,
including the Hamiltonian used,
the properties and construction of the two-body and three-body PESs,
and a more detailed description of the perturbative and full inclusion treatments.
In Sec.~\ref{sec:computation}, we provide a review of the computational techniques used in the simulations.
In Sec.~\ref{sec:results}, we present the results and a discussion thereof.
Finally, we present our conclusions and suggestions for future work.

\section{Model} \label{sec:model}
We model our system as a collection of $ N $ point-wise parahydrogen molecules
in three dimensions.
They are contained in an \textit{hcp} lattice within a rectangular prism,
with periodic boundary conditions applied in all directions.
The Hamiltonian for our system is
\begin{equation} \label{eq:hamiltonain_2b_3b}
    \hat{H} = - \frac{\hbar^2}{2m} \sum_{i=1}^{N} \nabla_i^2
              + \sum_{i < j}^{N}      V_2(r_{ij}) 
              + \sum_{i < j < k}^{N}  V_3(r_{ij}, r_{ik}, r_{jk}) ,
\end{equation}
\noindent
where
$ V_2 $ is the FSH pair interaction,
$ V_3 $ is the \abinitio three-body interaction,\cite{threebody:22ibra}
and $ r_{ij} $ is the relative distance between the centres of mass of molecules with labels $ i $ and $ j $.

The FSH potential is a static \abinitio pair interaction potential for parahydrogen.
In 2014, Faruk \etal applied the adiabatic hindered rotor (AHR) method\cite{hinderedrotor:10li, li2010analytic, dhe:12wang, wang2013new, li2013analytic, h2pes:13toby}
to a 6-dimensional \ce{H2}--\ce{H2} dimer potential
created by Hinde.\cite{ph2cluster:14faru, ph2cluster:15schm, h2pes:08hind}
This process produced several pointwise 1D pair potentials,
each of whose only parameter
is the relative distance between each molecule's centre of mass.
These potentials have been used to successfully simulate
and predict the experimentally observed first vibrational shifts of small clusters
of parahydrogen molecules.\cite{ph2cluster:14faru,ph2cluster:15schm,moleculeh2:20marr}
This AHR approximation has also recently been successfully applied
to help construct neural network interaction potentials.\cite{neural:22schr}

We refer to one of these aforementioned potentials,
that which gives the ground state interaction energy for two parahydrogen molecules,
as the FSH potential.
As a purely \abinitio pair potential,
the FSH potential does not implicitly incorporate higher many-body effects.
It has a both a deeper attractive well and a much more repulsive core than the SG and Buck potentials.
When used on its own,
it greatly overestimates the pressure of solid parahydrogen,
even at relatively low densities (below $ 0.04 \, \mathrm{\AA}^{-3} $).\cite{pathinteg:19ibra}

This research group recently published a static \abinitio three-body PES for parahydrogen.\cite{threebody:22ibra}
Its energies are calculated at the correlated coupled-cluster theory level, with single, double, and perturbative triple excitations. 
The calculations were performed using an augmented correlation-consistent triple zeta atom-centred basis set, supplemented by a ($3s3p2d$) midbond basis. 
The PES is isotropic.
Each hydrogen molecule is fixed to the vibrationally averaged ground state bond length of $ 1.449 $~\ANG.
The angular degrees of freedom are projected out using the $ L = 3 $ Lebedev quadrature,\cite{lebedev:76lebe}
in which each effective interaction energy is the average interaction energy of $ 27 $ space-fixed angular orientations.
The PES is constructed using the reproducing kernel Hilbert Space (RKHS) toolkit by Unke \etal\cite{rkhs:17unke}
The RKHS method is a machine learning method that has seen much success constructing multi-dimensional PESs.\cite{rkhs:01holl, rkhs:15zhai, rkhs:20kone, rkhs:21unke}
In addition, we apply a number of phenomenological modifications and extreme-range extensions to the PES.
These modifications ensure that the potential converges smoothly to the ATM potential at large intermolecular separations,
and that its energies increase in magnitude exponentially at short intermolecular separations.
The training data for the three-body potential only goes down to triangles with at least one side length greater than $ 2.2 $ \ANG.
This is very similar to the lower limit for the FSH potential.
Luckily, even at the highest densities relevant to the findings of this paper,
the parahydrogen molecules spend very little time at such short distances.

In principle, because parahydrogen is a boson,
we should treat it as an indistinguishable particle that obeys Bose statistics.
In the context of a PIMC simulation,
we account for Bose statistics using the worm algorithm,\cite{pathinteg:06bonia, pathinteg:06bonib}
which expresses the permutation of identical bosons
via the connection and exchange of beads between the ring polymers.
We find,
however,
that even at the highest densities explored here,
that bosonic exchanges between the molecules are almost entirely absent.
Thus in practice,
we can treat the parahydrogen molecules as distinguishable particles.

Inspired by previous work by Barnes and Hinde,\cite{pathinteg:17barn_1}
we will compare results from two different methods,
each of which incorporates the three-body interactions in a different way:
the perturbative method and the full inclusion method.
With no additional effort,
the perturbative treatment simulations also give us results
for the case where the three-body potential is used in neither sampling nor the estimators,
i.e. pure two-body potential simulations.
These methods are described in more detail below.

\subsection{Full Inclusion Method}
In the full incorporation method, we use the pair and three-body potentials in both the configuration sampling stage and the energy estimators.
During sampling, only three-body interactions between a given molecule and its 12 nearest neighbours are taken into account.
This is an appropriate approximation;
the three-body interaction falls off quickly enough that,
even at the highest densities considered in this paper,
the contribution from the pair potential beyond the nearest neighbours is typically 2 to 3 orders of magnitude greater than that from the three-body potential.
When using the energy estimator,
we follow the Attard minimum image convention\cite{threebody:91atta}
and include all three-body interactions in the lattice up to the cutoff distance of $ L / 2 $,
where $L$ is the shortest length of the simulation box.
The explicit inclusion of the three-body potential in the sampling stage greatly increases the computational costs.
The evaluation of the three-body potential itself is more expensive than for pair potentials,
and even when limited to the nearest-neighbours, there are 66 interactions to account for per molecule.
Compared to using pair interactions alone, the inclusion of the three-body interactions increases
the computational cost of the PIMC simulations by a factor of around 40.

\subsection{Perturbative Method}
In the perturbative method,
we use only the pair potential during configuration sampling,
but use both the pair and three-body potentials for the energy estimators.
Like in the full inclusion method, the energy estimators include all three-body interactions up to the cutoff distance $ L / 2 $.
Including the three-body potential perturbatively incurs a near negligible computational cost to the simulation.
This is because, relative to the sampling stage, the estimation stage takes up a very small portion of the simulation time.
To understand the accuracy trade-off of the perturbative method,
we should remember that the three-body potential is attractive at short ranges.
Its full inclusion in the sampling stage
makes high density configurations less energetically unfavourable,
and thus we should expect a change the positional distributions
of the molecules in the solid.
The accuracy of the perturbative method relies on the assumption that this change in positional distributions is very small.
Our findings suggest that this assumption is sound (see Secs.~\ref{sec:results:centroid} and \ref{sec:results:pairdist}).
This assumption also holds in solid helium-4.
Barnes and Hinde\cite{pathinteg:17barn_1} found that,
when compared to the perturbative method,
the full inclusion of the three-body interaction
only slightly shifted the pair distribution of solid helium-4 towards shorter distances.
This holds even up to the relatively high density of $ 0.15 \, \ANG^{-3} $.

In the remainder of this paper,
to describe the category of simulation that the presented results come from,
we will use the following labels.
Results from PIMC simulations with pair interactions only,
for both the sampling and the energy estimator,
are labelled as FSH-[2B].
Results from simulations with pair interactions and perturbatively incorporated three-body interactions
are labelled as FSH-[2B(3B)].
Results from simulations with pair interactions and fully incorporated three-body interactions
are labelled as FSH-[2B+3B].

\section{Computation} \label{sec:computation}
\subsection{Path Integral Monte Carlo}
Consider a system of particles described by a Hamiltonian $ \hat{H} $ at a finite temperature $ \beta = 1 / k_B T $.
The expectation value of the operator $ \hat{A} $ in this system can be described using
\begin{align} \label{eq:comp:expectation_of_operator}
    \langle \hat{A} \rangle
    &=
    \frac{1}{Z} \mathrm{Tr}
        \left\{
            \hat{A} e^{ - \beta \hat{H} }
        \right\} \nonumber \\
    &=
    \frac{1}{Z} \int \dd \vb{q} \dd \vb{q}^{\prime}
        \mel{ \vb{q}          }{ \hat{A}               }{ \vb{q}^{\prime} }
        \mel{ \vb{q}^{\prime} }{ e^{ - \beta \hat{H} } }{ \vb{q}          } ,
\end{align}
\noindent
where $ e^{ - \beta \hat{H} } $ is the density matrix,
$ Z = \mathrm{Tr} \left\{ e^{ - \beta \hat{H} } \right\} $ is the partition function,
and $ \vb{q} $ represents the positions of all particles in the system.

Using the path integral formulation we take the identity operator resolved onto the position space,
\begin{equation}
    \hat{I} = \sum \ket{\vb{q}} \bra{\vb{q}}
\end{equation}
\noindent
and apply it $ P $ times to discretize the integral Eq.~(\ref{eq:comp:expectation_of_operator}) in imaginary time
\begin{equation} \label{eq:comp:discretization}
    \mel{ \vb{q}^{\prime} }{ e^{ - \beta \hat{H} } }{ \vb{q} }
    =
    \int
    \prod_{i=2}^{P} \dd \vb{q}_i \times \prod_{i=1}^{P}
    \mel{ \vb{q}_{i} }{ e^{ - \tau \hat{H} } }{ \vb{q}_{i+1} } ,
\end{equation}
\noindent
where $ \tau = \beta / P $ is the imaginary time step,
and the index $ i $ labels the individual time steps.
We apply the boundary conditions $ \vb{q}_1 = \vb{q}^{\prime} $ and $ \vb{q}_{P+1} = \vb{q} $,
thus forming a ring of coordinates $ \{ \vb{q}_i \} $.
We can sample the integral Eq.~(\ref{eq:comp:expectation_of_operator}) with the PIMC method.\cite{pathinteg:95cepe}

\subsection{Trotter Error}
Note that when we separate
the Hamiltonian in Eq.~(\ref{eq:comp:discretization}) into potential and kinetic components,
we introduce a systematic Trotter factorization error into our expectation values.\cite{pathinteg:95cepe, pathinteg:17yan}
This error is a function of the imaginary time step size $ \tau $, and vanishes in the limit $ \tau \rightarrow 0 $.
A previous study of the EOS of solid parahydrogen shows that this Trotter factorization error is significant.\cite{pathinteg:19ibra}

One method to eliminate the Trotter error
is to perform several simulations,
each with a different value of $ \tau $.
Each simulation produces a different expectation value $ \langle \hat{A} \rangle_{\tau} $.
We then fit these expectation values to the equation\cite{pathinteg:17yan}
\begin{equation} \label{eq:comp:tau_extrapolation}
    \langle \hat{A} \rangle_{\tau}
    =
    \langle \hat{A} \rangle_{0}
    +
    b_2 \tau^2
    +
    b_4 \tau^4 \, ,
\end{equation}
\noindent
where $ \langle \hat{A} \rangle_{\tau} $ is the expectation value of $ \hat{A} $ in the limit $ \tau \rightarrow 0 $,
and $ b_2 $ and $ b_4 $ are fit parameters.
When both the kinetic and potential energies are small in magnitude,
the systematic Trotter errors are small,
and we can successfully use Eq.~(\ref{eq:comp:tau_extrapolation})
using relatively large values of $ \tau $ for fit values.
When the kinetic or potential energies are large,
we need smaller values of $ \tau $ to use Eq.~(\ref{eq:comp:tau_extrapolation}),
and this extrapolation method becomes less advantageous.

Another approach
is to select a small enough value of $ \tau $
such that the Trotter error becomes tolerably small.
This is much simpler,
but the resulting kinetic energy energies have a much larger variance.

We use both the ``small $\tau$'' and ``$\tau$ extrapolation'' methods,
each on a different range of densities,
to calculate different features of the EOS.

We use the ``small $\tau$'' method, using $ P = 960 $ beads
($ \tau \approx 2.48 \times 10^{-4} $ K$^{-1}$),
to calculate the energies in a sparse grid
between $ \rho = 0.024 \, \ANG^{-3} $ and $ \rho = 0.1 \, \ANG^{-3} $.
We find that, compared to the extrapolated value,
this choice of time step leads to a Trotter error in the total energy of
about $ 0.2 \, \% $ at $ 0.026 $ $\mathrm{\ANG}^{-3}$ and $ 0.04 $ $\mathrm{\ANG}^{-3}$,
and about $ 2.5 \, \% $ at $ 0.1 $ $\mathrm{\ANG}^{-3}$.
As we will see in later sections,
the simulation results at high densities are incredibly divorced from the experimental results,
and thus the increased systematic Trotter error at high densities should not be too concerning.
The standard error of the mean of the energies from the ``small $\tau$'' method
is on the order of $0.1$ cm$^{-1}$,
whereas the energies in concern span a range of nearly $ 4000 $ cm$^{-1}$
(see Sec.~\ref{sec:results:energy}).

We use the ``$ \tau $ extrapolation'' method
to calculate the energies in a fine grid
near the equilibrium density (around $ \rho_0 = 0.026 \, \ANG^{-3} $).
We perform multiple simulations at each density,
using $ P = \{ 64, 80, 96, 128, 192 \} $ beads,
and extrapolate the energies to $ \tau \rightarrow 0 $ using Eq.~(\ref{eq:comp:tau_extrapolation}).
The energies around the equilibrium density are low enough
for the extrapolation to work well,
and we need low variance results to accurately find the equilibrium density.
The standard error of the mean of the energies from the ``$\tau$ extrapolation'' method
is on the order of $0.01$ cm$^{-1}$,
whereas the energies around the equilibrium density span a range of only about $ 2 $ cm$^{-1}$.

\subsection{Finite System Size Error} \label{sec:comp:finite_size}
In systems with periodic boundary conditions,
we must apply minimum image rules to ensure the correct distances between molecules are chosen.
For example,
due to the periodicity,
two molecules near opposite edges of the simulation box are actually very close together,
and our simulations must account for this.
For the pair potential, we apply the standard minimum image convention with a cutoff distance of $ L / 2 $,
where $ L $ is the shortest length of the simulation box.
Interactions between pairs of molecules greater than $ L / 2 $ apart are ignored.
For three-body interactions,
this minimum image convention does not work -- it has a tendency to underestimate the size of triangles.
There are two common ways to overcome this.
Some authors use a cutoff distance of $ L/4 $.
However, at high densities, this distance becomes too short to give accurate results.\cite{threebody:13rio}
We instead apply the modifications to the minimum image convention developed by Attard,\cite{threebody:91atta}
which lets us keep all triangles for which each of its side lengths is less than the $ L / 2 $ cutoff.

Our simulations are performed using a finite number of molecules $ N $ inside a box of side length $ L $.
Thus the energy estimators cannot account for interactions with molecules outside of the simulation box,
and our potential energy expectation values suffer from systematic finite system size errors.
We correct these errors by applying potential tail corrections for both the two-body and three-body potentials.\cite{book:89alle}

\begin{figure} [H]
    \centering
    \includegraphics[width=\linewidth]{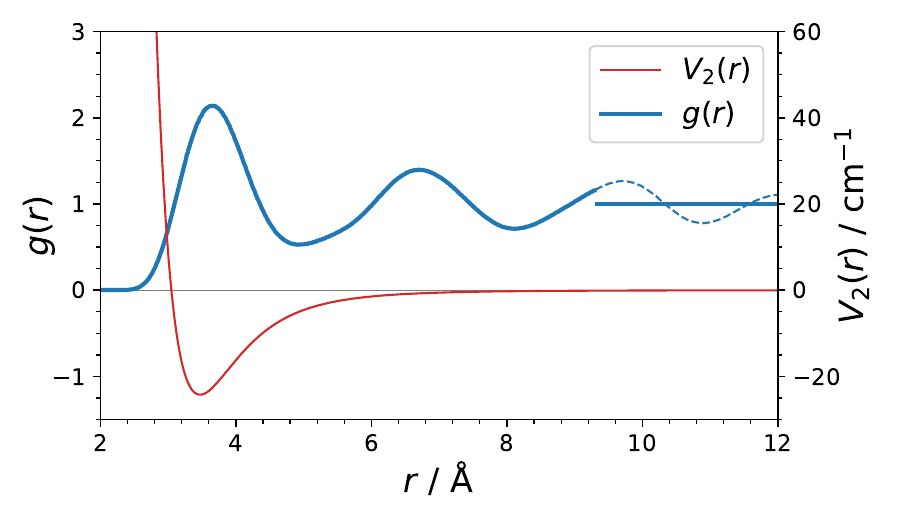}
    \caption{
        The radial pair distribution function $ g(r) $ (blue)
        from a simulation of $ N = 448 $ parahydrogen molecules
        using $ P = 64 $ beads,
        at a density of $ \rho = 0.026 \, \ANG^{-3} $.
        Also shown is the FSH pair potential $ V_2(r) $ (red).
        In the region $ r > L/2 \approx 9.3 \, \ANG $,
        the thin, dashed blue line is the $ g(r) $ of the $ N = 448 $ simulation,
        and the thick, horizontal blue line represents the tail correction
        made when we instead use a $ g(r) $
        from a simulation with $ N = 180 $ molecules.
    }
    \label{fig:paper2_fig0}
\end{figure}

To calculate the tail correction energy $ \epsilon^{(t)}_{N} $ for the pair potential,
we assume that the radial pair distribution function $ g(r) $
outside our simulation box is uniform (equal to unity),
and then integrate the pair potential outside this box.
\begin{align}
    \epsilon^{(t)}_N
    &=
    \frac{N \rho}{2} \int_{L/2}^{\infty}
        \dd[3]r \ V_2(r) g(r) \phantom{aaaaa} \mathrm{(pair)} \nonumber \\
    &\approx
    \frac{N \rho}{2} \int_{L/2}^{\infty}
        \dd[3]r \ V_2(r) \label{eq:pair_tail_correction}
\end{align}
\noindent
In Fig.~\ref{fig:paper2_fig0},
we show the radial pair distribution function $ g(r) $
at the equilibrium density of $ \rho = 0.026 \, \ANG^{-3} $,
from a simulation of $ N = 448 $ parahydrogen molecules,
using $ P = 64 $ beads.
We also show the FSH pair potential $ V_2(r) $.
For $ N = 448 $ at this density,
the cutoff distance is roughly $ 12.3 \, \ANG $.
The radial pair distribution function for a
simulation with $ N = 180 $ molecules
is effectively identical to its shown counterpart with $ N = 448 $ molecules,
but only until its own shorter cutoff distance of $ L/2 \approx 9.3 \, \ANG $.
The thin, oscillating, dashed blue line
beyond the cutoff distance $ L/2 \approx 9.3 \, \ANG $ in Fig.~\ref{fig:paper2_fig0}
is the ``tail'' that we want to approximate.
When we set $ g(r) = 1 $ for $ r > L/2 $ in Eq.~(\ref{eq:pair_tail_correction}),
we can represent this visually in Fig.~\ref{fig:paper2_fig0},
by replacing the ``true tail'',
(the thin, oscillating blue line),
with the thick, horizontal blue line that overlaps it.

The tail correction is the most effective
when performed at distances for which
the oscillations in the radial pair distribution function,
and the magnitude of the pair potential,
are both relatively small.
For example,
at the equilibrium density,
for a simulation with $ N = 180 $ molecules,
we see in Fig.~\ref{fig:paper2_fig0} that the tail correction
would begin well into the $ r^{-6} $ decay of the pair interaction.
However, we will soon see that the tail correction is still reasonable
even at the highest densities explored in this paper,
especially when compared with the total energy scales of the calculations.

For the three-body interaction tail correction we follow Eq.~(2.150) in Ref.~[\onlinecite{book:89alle}],
and account for interactions from all triangles with at least one side length greater than $ L / 2 $.
This calculation requires the three-body distribution function,
which,
following the aforementioned reference,
we approximate using a product of three pair distribution functions $ g(r) $ from the simulation.
\begin{figure} [H]
    \centering
    \includegraphics[width=\linewidth]{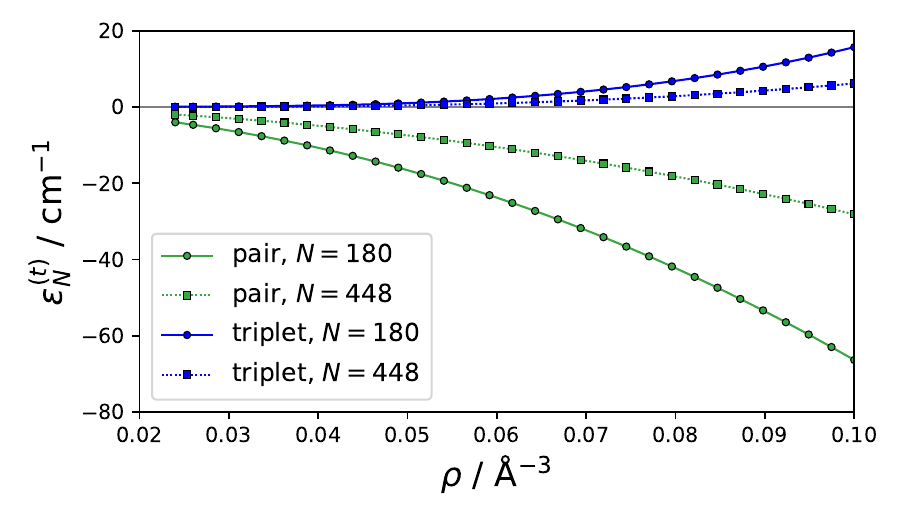}
    \caption{
        The potential tail correction energy per molecule $ \epsilon^{(t)}_N $ as a function of density $ \rho $.
        These energies are added to the simulation energies to correct for
        the finite system size error.
        Shown are the correction energies for simulations
            of  $ N = 180 $ particles (solid, circles)
            and $ N = 448 $ particles (dotted, squares),
        separated into the pair interactions (green)
        and the three-body interactions (blue).
    }
    \label{fig:paper2_fig1}
\end{figure}

To test how well the tail corrections work,
we perform perturbative treatment simulations of
solid parahydrogen for two system sizes, $ N = 180 $ and $ N = 448 $.
These simulations take place at temperature $ T = 4.2 $ K, using $ P = 64 $ time slices,
and for $ 31 $ density values between $ 0.024 $ \ANG$^{-3}$ and $ 0.1 $ \ANG$^{-3}$.
In Fig.~\ref{fig:paper2_fig1},
we show the pair and three-body energy corrections for both lattice sizes.
As expected, the tail corrections are smaller in magnitude for the larger $ N = 448 $ case than for the smaller $ N = 180 $ case.
This is expected, because a larger system should have a smaller system size error to correct for.
At all considered densities, the pair tail correction is attractive, while the three-body tail correction is repulsive.
In addition, the pair tail correction is always greater in magnitude than that of the three-body tail correction.
This reflects the fact that,
at large distances,
the pair potential decays with an attractive $ - C_6 r^{-6} $ interaction energy,
while the three-body potential decays with a weaker, repulsive $ C_9 r^{-9} $ interaction energy.

\begin{figure} [H]
    \centering
    \includegraphics[width=\linewidth]{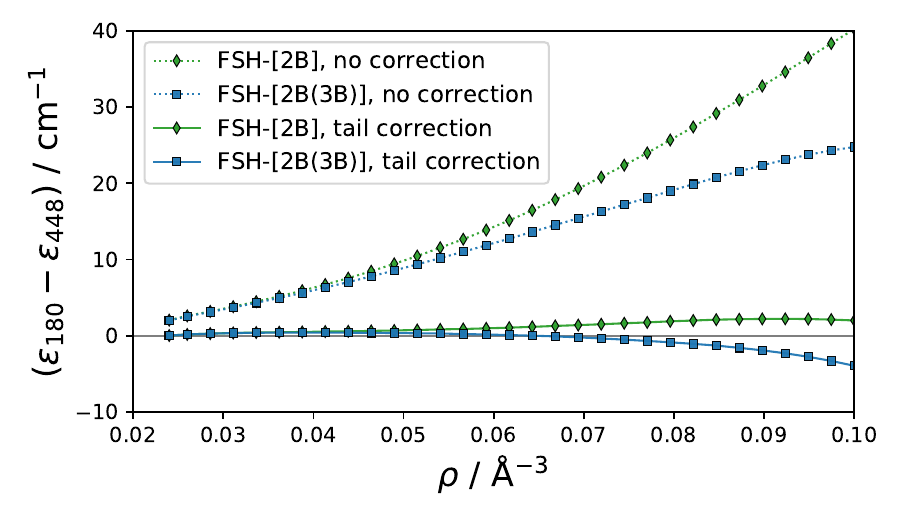}
    \caption{
        The difference in the energy per molecule
        calculated from simulations using
        $ N = 180 $ particles ($ \epsilon_{180}$)
        and $ N = 448 $ particles ($ \epsilon_{448}$).
        The differences are calculated for both
        the FSH-[2B] energies (green, diamonds)
        and the FSH-[2B(3B)] energies (blue, squares).
        The solid curves represent the cases
        where simulations of both sizes have had their
        respective potential tail corrections applied before the difference is taken
        (\textit{i.e.} $ \epsilon_N \approx \epsilon^{(s)}_N + \epsilon^{(t)}_N $).
        The dotted curves represent the cases
        where the tail corrections have been ignored
        (\textit{i.e.} $ \epsilon_N \approx \epsilon^{(s)}_N $).
        For the FSH-[2B] case,
        the tail correction $ \epsilon^{(t)}_N $ is made up of only the pair tail contribution
        in Fig.~\protect\ref{fig:paper2_fig1},
        whereas for the FSH-[2B(3B)] case,
        $ \epsilon^{(t)}_N $ is made up of both the pair and three-body tail contributions.
        Having an energy difference close to zero
        represents a successful finite system size correction.
    }
    \label{fig:paper2_fig2}
\end{figure}

From the simulations we recover
the average energy per particle for a system of $ N $ particles, $ \epsilon^{(s)}_N $,
for which the tail correction has not been applied.
We now consider two cases.
We can calculate the total energy per particle either with the tail correction,
\textit{i.e.} $ \epsilon_N \approx \epsilon^{(s)}_N + \epsilon^{(t)}_N $,
or without the tail correction,
\textit{i.e.} $ \epsilon_N \approx \epsilon^{(s)}_N $.
We then calculate the difference in the energy per particle of both system sizes,
$ \epsilon_{180} - \epsilon_{448} $,
for both the corrected and uncorrected cases.
If the tail correction is effective,
we should expect that it makes up for the difference in the simulation sizes,
and thus $ \epsilon_{180} - \epsilon_{448} $ should be close to zero.
We can verify this in Fig.~\ref{fig:paper2_fig2}.
Without the tail correction, the energy gap per particle $ \epsilon_{180} - \epsilon_{448} $ grows with the density.
The application of the tail correction suppresses the energy difference almost completely at lower densities,
and still relatively well at the highest densities considered.
Thus with the tail corrections we can justify performing simulations using only $ N = 180 $ particles.

\section{Results} \label{sec:results}
\subsection{Energy-Density Equation of State} \label{sec:results:energy}
We perform PIMC simulations of solid parahydrogen at $ T = 4.2 $ K,
between densities of $ 0.024 \, \ANG^{-3} $ and $ 0.1 \, \ANG^{-3} $ for the perturbative simulations, and
between densities of $ 0.0617 \, \ANG^{-3} $ and $ 0.1 \, \ANG^{-3} $ for the full inclusion simulations.
Due to the expense of the full inclusion calculations, they are only carried out at the higher densities,
where the influence of the three-body potential is the greatest.
We use the standard potential energy estimators for both the pair and three-body potential energies,
taking into account periodic boundary conditions and potential tail corrections,
as discussed in Sec.~\ref{sec:computation}.
We use the primitive estimator for the kinetic energy.
\begin{figure} [H]
    \centering
    \includegraphics[width=\linewidth]{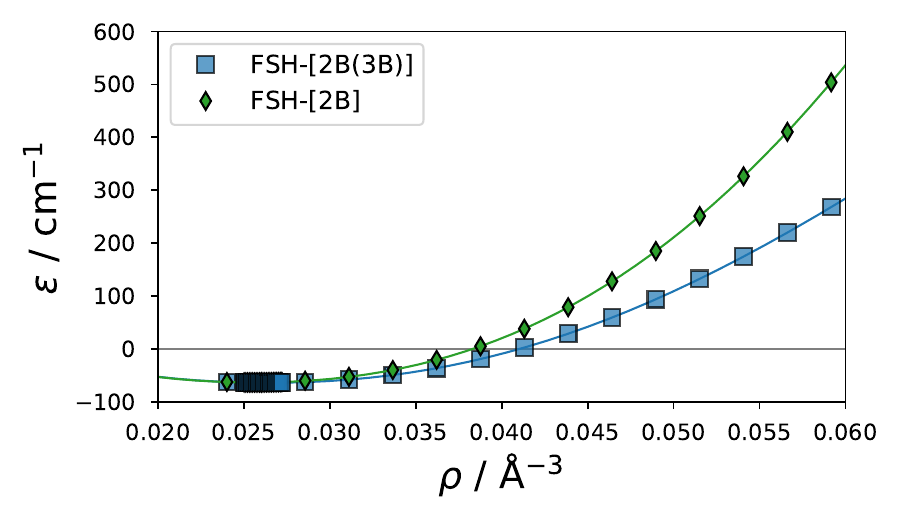}
    \includegraphics[width=\linewidth]{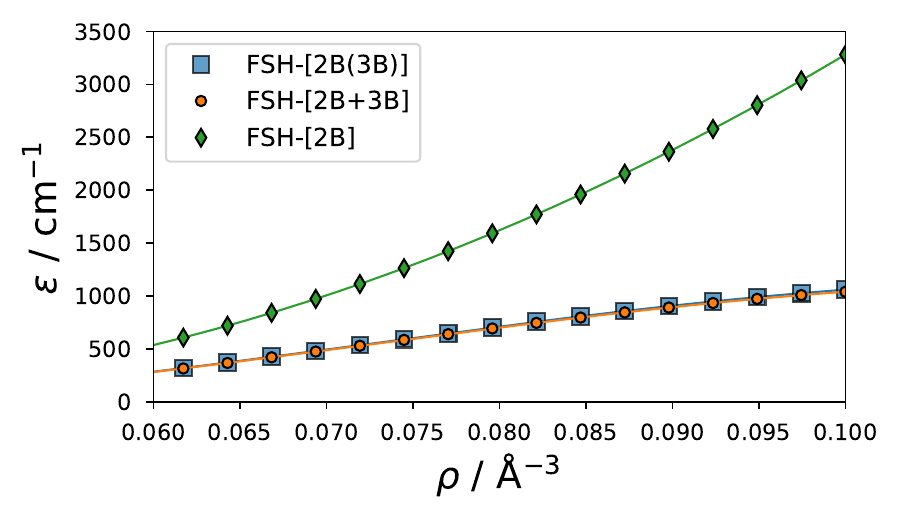}
    \caption{
        The energy per molecule $\epsilon$
        as a function of the density $\rho$ of solid parahydrogen.
        The data points are shown for simulations performed under
        the FSH-[2B] conditions (green, diamonds),
        the FSH-[2B(3B)] conditions (blue, squares),
        and the FSH-[2B+3B] conditions (orange, circles).
        The curves represent the fitting of the data to a modified Birch equation,
        Eq.~(\protect\ref{eq:results:energy:birch_fit}).
        All energies were calculated using the ``small $\tau$'' method
        to account for the Trotter error,
        except for those in the densely packed region around $ \rho_0 = 0.026 \, \ANG^{-3} $,
        which were calculated using the ``$\tau$ extrapolation'' method.
    }
    \label{fig:fig3}
\end{figure}

We fit the FSH-[2B] and FSH-[2B(3B)] total energies per particle to a modified Birch EOS\cite{ph2solidexp:00cohe}
\begin{equation} \label{eq:results:energy:birch_fit}
    \epsilon(\rho)
    =
    \epsilon_0
    -
    P_0 / \rho
    +
    \sum_{n=1}^{5} \tilde{\kappa}_n \rho^{2n/3} \ .
\end{equation}
The fit parameters $ \epsilon_0 $, $ P_0 $, and $ \{ \tilde{\kappa}_n \} $
are shown in Table~\ref{tab:results:energy:birch_fit_parameters}.
For the purposes of being able to create a curve
for the pressure-density figures in Sec.~\ref{sec:results:pressure},
we also fit the FSH-[2B+3B] total energies to Eq.~(\ref{eq:results:energy:birch_fit}).
However,
because the energies for the FSH-[2B+3B] simulations
were collected for a more limited range of densities,
the fit parameters are very highly correlated to one another
and are not very meaningful.

\begin{table*} [ht]
    \caption{
        Fit parameters for the energy-density EOS,
        Eq.~(\protect\ref{eq:results:energy:birch_fit}),
        for the FSH-[2B] and FSH-[2B(3B)] simulation sets.
    }
    \begin{ruledtabular}
	    \label{tab:results:energy:birch_fit_parameters}
        \begin{tabular}{lcc}
            $                                          $ &   \mbox{FSH-[2B]}           &   \mbox{FSH-[2B(3B)]}        \\
            \hline
            $ \epsilon_0 (\mathrm{cm}^{-1}          )$ & $  16.78      \pm 0.87      $ & $  18.67      \pm 0.78     $ \\
            $ P_0        (\mathrm{cm}^{-1} \ANG^{-3})$ & $ -4842.4     \pm 222.7     $ & $ -5763.3     \pm 200.8    $ \\
            $ \kappa_1   (\mathrm{cm}^{-1} \ANG^{2} )$ & $  118465.1   \pm 5070.8    $ & $  149565.7   \pm 4572.1   $ \\
            $ \kappa_2   (\mathrm{cm}^{-1} \ANG^{4} )$ & $ -1312002.4  \pm 53696.9   $ & $ -1789644.0  \pm 48415.6  $ \\
            $ \kappa_3   (\mathrm{cm}^{-1} \ANG^{6} )$ & $  6880148.8  \pm 302556.8  $ & $  10629544.1 \pm 272798.6 $ \\
            $ \kappa_4   (\mathrm{cm}^{-1} \ANG^{8} )$ & $ -14867506.4 \pm 877135.3  $ & $ -28398452.1 \pm 790863.2 $ \\
            $ \kappa_5   (\mathrm{cm}^{-1} \ANG^{10})$ & $  14135049.1 \pm 1030210.7 $ & $  26646661.8 \pm 928881.8 $ \\
        \end{tabular}
    \end{ruledtabular}
\end{table*}

In Fig.~\ref{fig:fig3},
we plot the energy-density results for the FSH-[2B], FSH-[2B(3B)], and FSH-[2B+3B] simulation cases.
The FSH-[2B] energies increase with density far more strongly than the other categories.
The FSH-[2B(3B)] and FSH-[2B+3B] energies are very similar to one another,
with the latter only slightly lower than the former.
At $ 0.0617 \, \AA^{-3} $ their difference is $ 3.6 \, \mathrm{cm}^{-1} $ (about $ 1 \, \% $ error),
and even up to $ 0.1 \, \AA^{-3} $, the difference is only $ 18.1 \, \mathrm{cm}^{-1} $ ($ 1.8 \, \% $).

\begin{figure} [H]
    \centering
    \includegraphics[width=\linewidth]{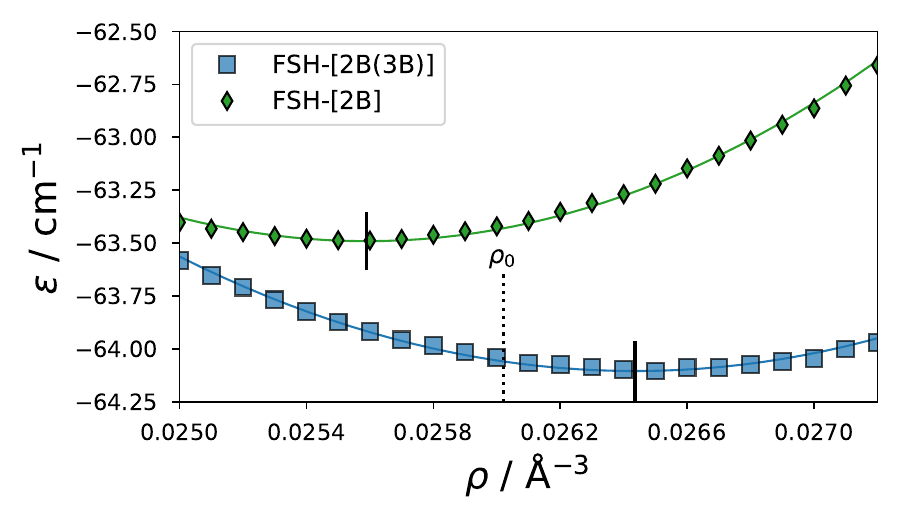}
    \caption{
        The energy per molecule $\epsilon$ as a function of density $\rho$,
        for simulations performed under
        the FSH-[2B] conditions (green, diamonds),
        and the FSH-[2B(3B)] conditions (blue, squares).
        The minima of the FSH-[2B] and FSH-[2B(3B)] curves
        are indicated by the thin and thick solid vertical black lines, respectively.
        The experimental equilibrium density $\rho_0$
        is indicated by the vertical dotted line.
        The energies in this figure were calculated using the ``$\tau$ extrapolation'' method
        to account for the Trotter error.
    }
    \label{fig:fig3_equil}
\end{figure}

The equilibrium density of the EOS
is indicated by the minimum of the energy-density curve.
In Fig.~\ref{fig:fig3_equil},
we show the FSH-[2B] and FSH-[2B(3B)] energy curves
near the experimental equilibrium density $ \rho_0 = 0.0260 \, \ANG^{-3} $.\cite{ph2solidexp:80silv}
Our comparisons of the FSH-[2B(3B)] and FSH-[2B+3B],
as well as later results shown in Fig.~\ref{fig:paper2_fig6},
show that the difference between the perturbative and full inclusion energies
rapidly decreases as we go to lower densities.
Thus we expect the equilibrium density to be low enough
such that the perturbative and full inclusion results have similar minima.
The FSH-[2B] curve underpredicts the minimum, at $ \rho = 0.02551 \, \ANG^{-3} $,
reproducing our findings in Ref.~[\onlinecite{pathinteg:19ibra}].
Meanwhile,
the FSH-[2B(3B)] curve overpredicts the minimum, at $ \rho = 0.02646 \, \ANG^{-3} $.
This indicates that,
when the three-body interaction is used as the sole many-body correction term,
its attractiveness overcorrects the repulsiveness of the pair interaction
even as low as the equilibrium density.

\begin{figure} [H]
    \centering
    \includegraphics[width=\linewidth]{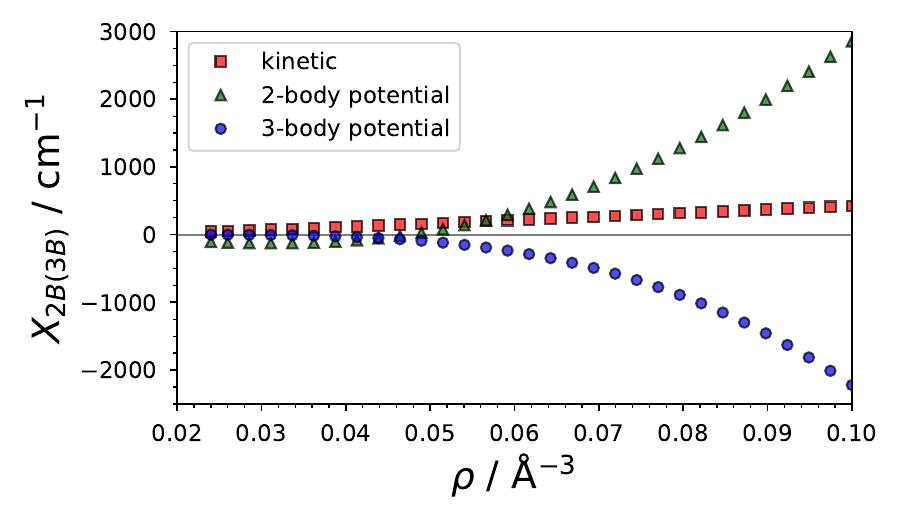}
    \caption{
        The energy per particle as a function of density $\rho$,
        separated into its
        kinetic (red, squares),
        pair potential (green, triangles),
        and three-body potential (blue, circles)
        components.
        The simulations were carried out under the FSH-[2B(3B)] conditions.
        The variable $X$ on the vertical axis label
        is a placeholder for the various energy components.
    }
    \label{fig:paper2_fig5}
\end{figure}

In Fig.~\ref{fig:paper2_fig5},
we show the individual energies that comprise the FSH-[2B(3B)] total energy curve,
i.e. the kinetic, the pair potential, and the three-body potential energies.
At low densities, all three types of energies have comparable magnitudes.
As the density increases,
both the pair and three-body potential energies grow exponentially in magnitude,
each at roughly the same rate.
The kinetic energy is much smaller in magnitude, and increases linearly.
However, the findings of Omiyinka and Boninsegni\cite{h2pes:13omiy}
suggest that this linear trend only holds for lower densities.
It appears clear that the potential energies
play the most significant role in the EOS of solid parahydrogen.

Additionally,
we can see how the inclusion or omission of the three-body interaction during sampling
affects each of the aforementioned three energies.
In Fig.~\ref{fig:paper2_fig6},
we take the kinetic, pair potential, and three-body potential energies
calculated using both the full inclusion and perturbative simulations,
and plot their respective differences as a function of density
between $ \rho = 0.06 \, \ANG^{-3} $ and $ \rho = 0.1 \, \ANG^{-3} $.
The full inclusion of the three-body interaction
causes a shift in energy for all three energy types,
and the magnitude of this shift increases with density.
These results give some insight as to why the total energy
from the perturbative and full inclusion simulations are so similar.
The increase in the pair potential energy
nearly counteracts
the combined decrease in the kinetic and three-body potential energies.
Our findings are in line with those of Barnes and Hinde in their simulations of solid helium,\cite{pathinteg:17barn_1}
wherein the perturbative and full inclusion treatments of the helium three-body potential of Cencek \etal\cite{he4pes:09cenc}
give very similar total energies.

\begin{figure} [H]
    \centering
    \includegraphics[width=\linewidth]{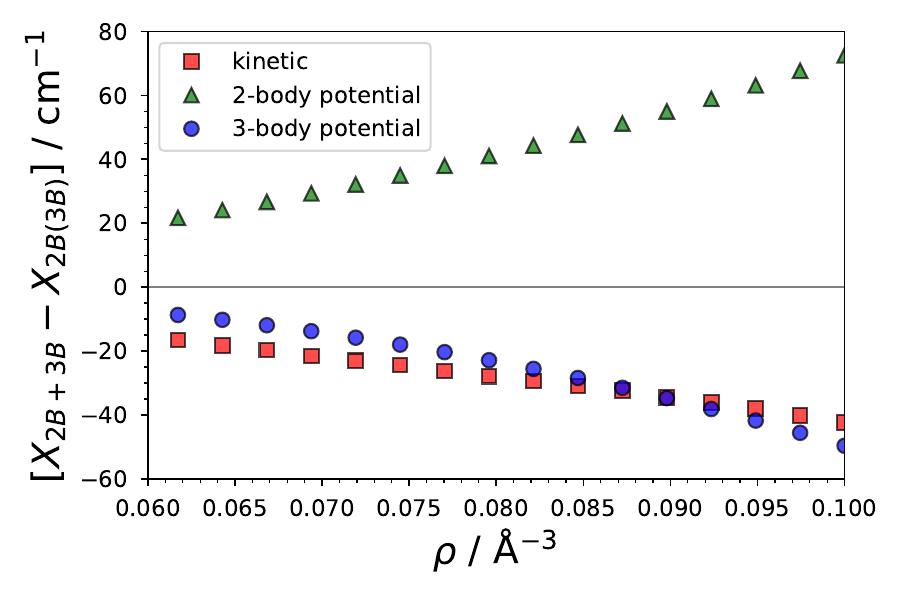}
    \caption{
        The difference in energy components as a function of density $\rho$,
        between energies calculated from FSH-[2B+3B] and FSH-[2B(3B)] simulations.
        The component labels match those
        in Fig.~\protect\ref{fig:paper2_fig5}.
        The variable $X$ on the vertical axis label
        is a placeholder for the various energy components.
    }
    \label{fig:paper2_fig6}
\end{figure}

To a large degree of accuracy,
when studying properties that depend only on the total energy,
it is enough for us to perform our simulations with only a perturbative treatment of the three-body potential.
However, if we need the individual kinetic and potential energies of the solid,
then the three-body potential must be included in the sampling.

\subsection{Pressure-Density Equation of State} \label{sec:results:pressure}
\begin{figure} [H]
    \centering
    \includegraphics[width=\linewidth]{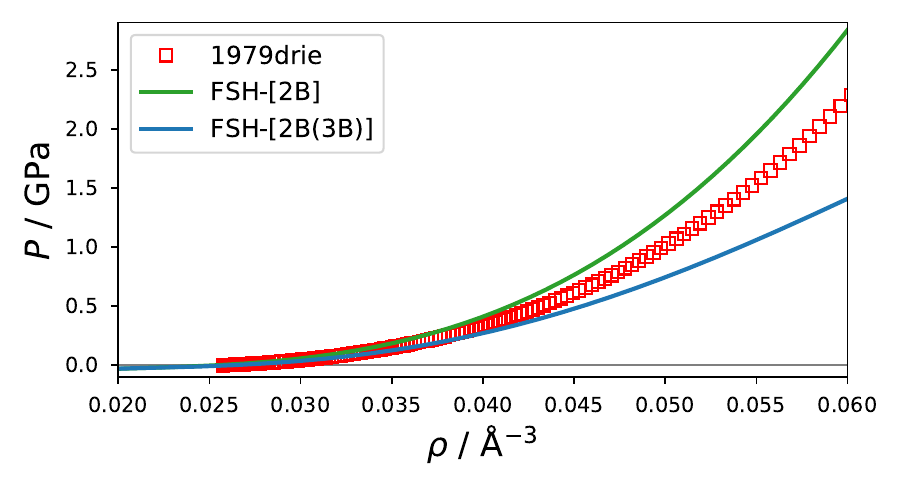}
    \includegraphics[width=\linewidth]{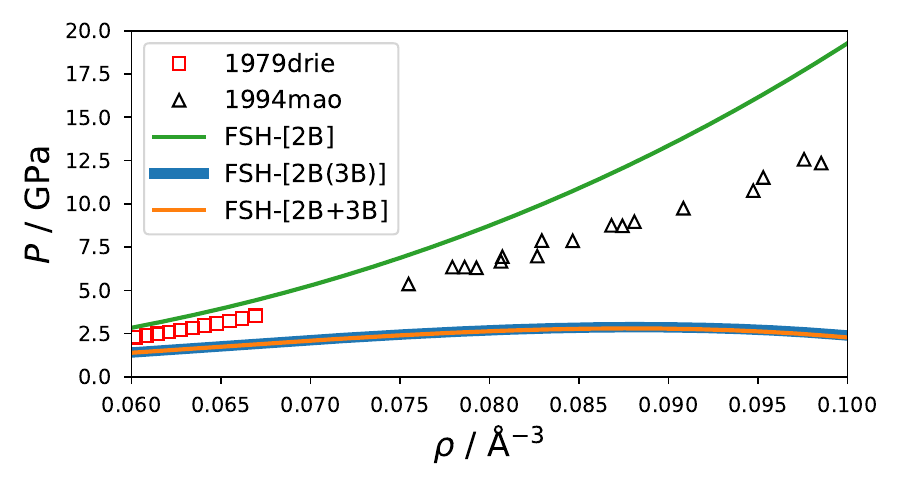}
    \caption{
        The pressure $P$ of solid parahydrogen
        as a function of density $\rho$.
        The curves are calculated using
        Eq.~(\protect\ref{eq:results:pressure:pressure_from_energy}),
        and correspond to simulations performed under
        the FSH-[2B] conditions (green),
        the FSH-[2B(3B)] conditions (blue),
        and the FSH-[2B+3B] conditions (orange).
        Also shown are experimental measurements
        from Ref.~[\onlinecite{ph2solidexp:79drie}] (red squares)
        and Ref.~[\onlinecite{ph2solidexp:94mao}] (black triangles).
    }
    \label{fig:fig7}
\end{figure}

We calculate pressures from the energy curves in Sec.~\ref{sec:results:energy} using
\begin{equation} \label{eq:results:pressure:pressure_from_energy}
    P(\rho)
    = \rho^2 \pdv{\epsilon}{\rho} \eval_{T}
    = P_0 + \frac{2}{3} \sum_{n=1}^{5} n \tilde{\kappa}_n \rho^{(2n+3)/3}
\end{equation}
\noindent
In Fig.~\ref{fig:fig7},
we show the pressure-density curves
for the FSH-[2B], FSH-[2B(3B)], and FSH-[2B+3B] cases
alongside experimental measurements. 

As expected from Ref.~[\onlinecite{pathinteg:19ibra}],
the FSH-[2B] curve greatly overshoots experiment,
even near the saturation density.
Meanwhile, the FSH-[2B(3B)] curve significantly underestimates the experimental data,
even at moderately low densities (below $ 0.04 \, \ANG^{-3} $).
As the density increases further,
the FSH-[2B(3B)] provides an increasingly worsening prediction
of the experiment than the FSH-[2B] curve.
Beyond $ 0.08 \, \ANG^{-3} $,
the three-body potential even leads to an unphysical decrease in the pressure-density curve.
The combination of the pair and three-body interaction potentials
is far too attractive at high densities.

Similar results have been seen in prior simulation studies of solid helium.
Barnes and Hinde\cite{pathinteg:17barn_1}
carried out PIGS simulations of solid helium-4
using the Aziz pair potential\cite{he4pes:87aziz}
and the three-body Cencek potential.\cite{he4pes:09cenc}
They found that this combination of interactions
leads to an underestimation of the pressure
when the helium is compressed to sufficiently low molar volumes.
Chang and Boninsegni\cite{pathinteg:01chan}
performed PIMC simulations of liquid and solid helium-4
using two separate combinations
of pair and three-body interactions.\cite{he4pes:79aziz, hepes3b:73bruc, he4pes:97janz, hepes3b:96cohe}
For both combinations of potentials,
their simulations underestimated the pressure curve
of solid helium-4 at low molar volumes.

We should note that many assumptions in our model of solid parahydrogen,
for example,
that it is an \textit{hcp} lattice,
and that it is composed of freely rotating molecules,
depend on solid parahydrogen being in the low-pressure phase I form.\cite{ph2solidexp:07pick}
At low temperatures,
solid parahydrogen remains in its phase I form below about $ 60 $ GPa,\cite{ph2solidexp:20greg}
which is much greater than what any of the pressure curves
in Fig.~\ref{fig:fig7} reach.
Thus, we can be confident that
in all simulations done for this paper,
the solid is in phase I.

\subsection{Centroid Distribution} \label{sec:results:centroid}
\begin{figure} [H]
    \centering
    \includegraphics[width=\linewidth]{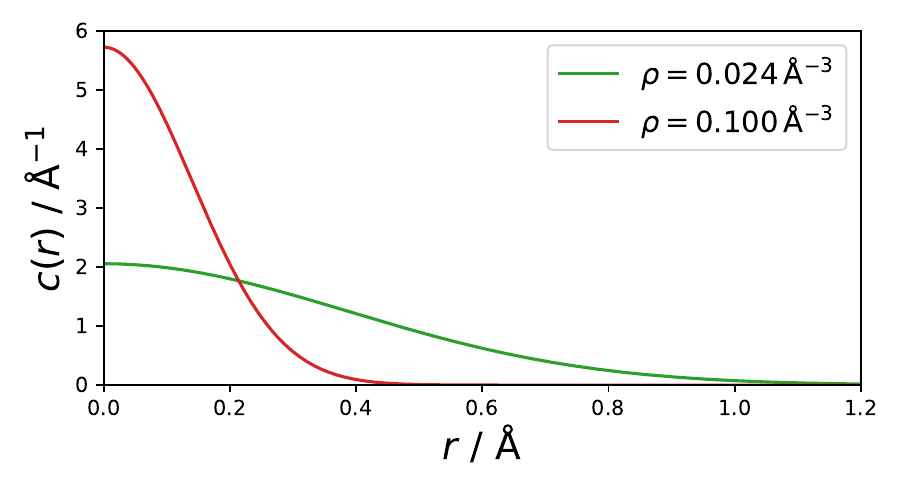}
    \caption{
        Normalized distribution of bead positions
        from the centroid of a parahydrogen molecule in solid parahydrogen,
        fit to Eq.~(\protect\ref{eq:results:centroid:gaussian_fit}).
        The results are collected from an FSH-[2B(3B)] simulation
        at a relatively low density ($\rho = 0.024 \, \ANG^{-3}$, green curve)
        and a relatively high density ($\rho = 0.1 \, \ANG^{-3}$, burgandy curve).
        The distribution is sharper and more classical at the higher density.
    }
    \label{fig:paper2_fig8}
\end{figure}

Solid parahydrogen is a quantum solid,
and its molecules have a large zero-point motion about their nominal lattice sites.
This behaviour is largely a consequence of
the low mass of, and weak intermolecular interactions between, the parahydrogen molecules.
The PIMC method accounts for this motion
by representing the quantum ``spreading'' of a particle in space
using the position distribution of beads in the PIMC ring polymer.
We can observe the zero-point motion of the molecules
by sampling the distance of the beads from the centroid of their corresponding molecule
\begin{equation} \label{eq:results:centroid:centroid_estimator}
    \hat{c}(r) =
    \Bigg\langle
        \sum_{n = 1}^{N} \sum_{i = 1}^{P}
        \delta( r - \left| \vb{r}_{i, n} - \vb{r}_{c, n} \right| )
    \Bigg\rangle
\end{equation}
\noindent
where $ \vb{r}_{i, n} $ is the $ i^{\rm th} $ bead on the $ n^{\rm th} $ particle,
and $ \vb{r}_{c, n} $ is the centroid of the $ n^{\rm th} $ particle, given by
\begin{equation} \label{eq:results:centroid:particle_centroid}
    \vb{r}_{c, n} =
    \frac{1}{P} \sum_{i = 1}^{P}
    \vb{r}_{i, n} \ .
\end{equation}
\noindent
The above estimator $ \hat{c}(r) $ produces a histogram in the shape of a Gaussian multiplied by $ r^2 $.
We rescale $ \hat{c}(r) $, and normalize it from $ r = [0, \infty) $ to get
\begin{equation} \label{eq:results:centroid:gaussian_fit}
    c(r) = \sqrt{\frac{2}{\pi}} \frac{1}{\sigma} \exp \left[ -\frac{1}{2} \frac{r^2}{\sigma^2} \right] .
\end{equation}
Equation.~(\ref{eq:results:centroid:gaussian_fit}) allows us to calculate the standard deviation $ \sigma $,
which represents the spread of the position distribution of each parahydrogen molecule.
We show examples of Gaussian distributions representing this zero-point motion in Fig.~\ref{fig:paper2_fig8},
for two different densities.

\begin{figure} [H]
    \centering
    \includegraphics[width=\linewidth]{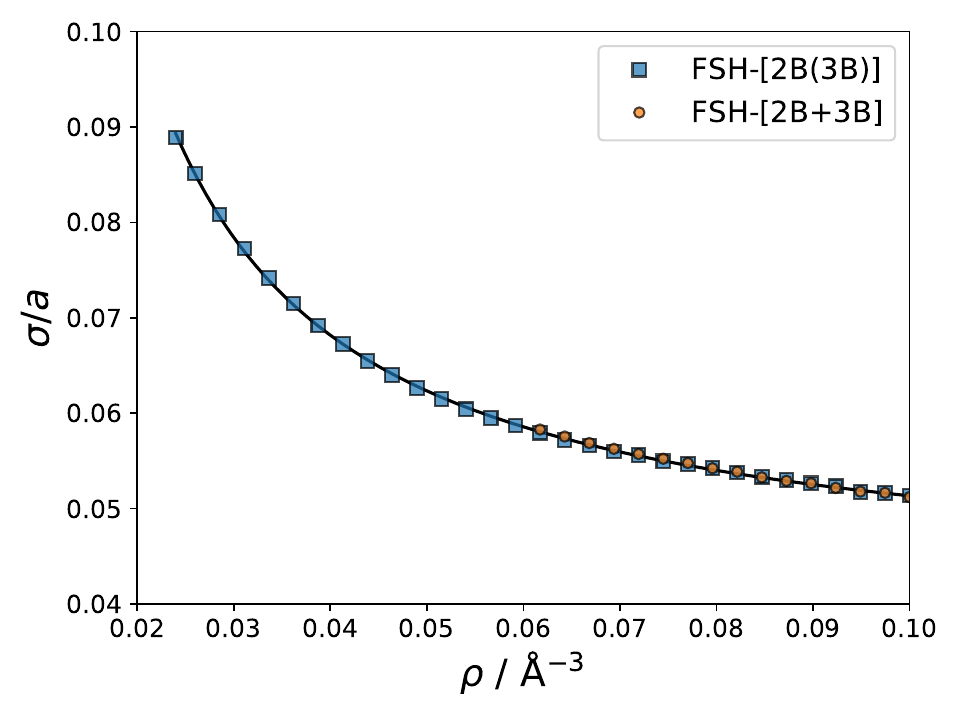}
    \caption{
        The ratio $\sigma/a$ as a function of density $\rho$,
        collected from simulations performed under
        the FSH-[2B(3B)] conditions (blue, squares),
        and the FSH-[2B+3B] conditions (orange, circles).
        Here,
        $\sigma$ is the standard deviation of the centroid bead distribution
        in Eq.~(\protect\ref{eq:results:centroid:gaussian_fit}),
        and $a$ is the lattice constant of the solid at that density.
        The curve represents the fit
        to Eq.~(\protect\ref{eq:results:centroid:ratio_fit}).
        The perturbative treatment of the three-body interaction has no effect
        on the structure of the solid,
        and so the FSH-[2B] results would be identical to the FSH-[2B(3B)] results.
    }
    \label{fig:paper2_fig9}
\end{figure}

In Fig.~\ref{fig:paper2_fig9},
we show the ratio $ \sigma / a $ as a function of density $ \rho $,
collected from perturbative and full inclusion treatment simulations.
Here, $ a $ is the lattice constant.
We see that the value of $ \sigma $ decreases with density.
As mentioned earlier,
one of the contributors to the presence of zero-point motion
is the weak intermolecular interaction between parahydrogen molecules.
As the density increases, the repulsive interactions become stronger,
and reduce the spread of the bead positions about their nominal lattice sites.
The structure becomes sharper and more classical at higher densities.

Even at the highest densities considered,
where the effect of the three-body potential is the strongest,
there is negligible difference in the centroid distributions between the two cases.
Thus it appears that,
despite the fact that the addition of the three-body potential
increases the average distance between neighbouring beads
(as indicated by the decrease in the kinetic energy seen in Sec.~\ref{sec:results:energy}),
the overall quantum mechanical ``spreading'' of each individual particle remains roughly unchanged.

We can fit the perturbative treatment data shown in Fig.~\ref{fig:paper2_fig9} to the fit curve
\begin{equation} \label{eq:results:centroid:ratio_fit}
    \frac{\sigma}{a}
    =
    s_0 + \frac{s_1}{\rho} + \frac{s_2}{\rho^2} ,
\end{equation}
\noindent
where $ s_0 = 4.63(2) \times 10^{-2} $, $ s_1 = 1.08(2) \times 10^{-3} \, \ANG^{-3} $, and $ s_2 = 5.1(5) \times 10^{-6} \, \ANG^{-6} $.

\subsection{Pair Distribution Functions} \label{sec:results:pairdist}
The inclusion of the three-body potential softens the overall interaction energy at high densities,
leading to changes in the structural properties of the lattice.
Two quantities we can look at to see this change
are the radial pair distribution function $ g(r) $,
whose (un-normalized) estimator is given by
\begin{equation} \label{eq:results:pairdist:radial_pairdist}
    \hat{g}(r) =
    \Bigg\langle
        \sum_{n<m}^N \sum_i^P
        \delta ( r - \left| \vb{r}_{i, n} - \vb{r}_{i, m} \right| )
    \Bigg\rangle
\end{equation}
\noindent
and the centroid radial pair distribution function $ g_c(r_c) $,
whose (un-normalized) estimator is
\begin{equation} \label{eq:results:pairdist:centroid_pairdist}
    \hat{g_c}(r_c) =
    \Bigg\langle
        \sum_{n<m}^N
        \delta ( r_c - \left| \vb{r}_{c, n} - \vb{r}_{c, m} \right| )
    \Bigg\rangle
\end{equation}
\noindent
where $ \vb{r}_{c, n} $ is given by Eq.~(\ref{eq:results:centroid:particle_centroid}).

\begin{figure} [H]
    \centering
    \includegraphics[width=\linewidth]{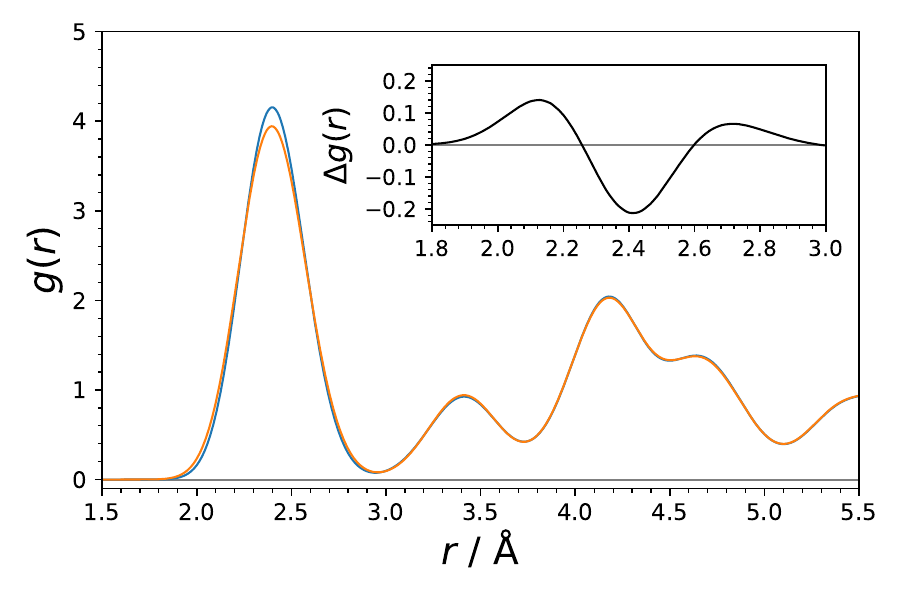}
    \caption{
        The radial pair distribution function $g(r)$
        of solid parahydrogen at a density $\rho = 0.1 \, \ANG^{-3}$,
        given by Eq.~(\protect\ref{eq:results:pairdist:radial_pairdist}).
        The results are collected from simulations performed under
        the FSH-[2B(3B)] conditions (blue, underlying curve),
        and the FSH-[2B+3B] conditions (orange, overlaying curve).
        Readers should focus on regions where the underlying blue curve is visible.
        In the inset we show $\Delta g(r)$,
        which is the difference between the orange and blue curves
        in the region around the first shell of the two $g(r)$ distributions.
    }
    \label{fig:paper2_fig10}
\end{figure}

In Fig.~\ref{fig:paper2_fig10},
we show an example of the normalized $ g(r) $ curves
collected from perturbative and full inclusion treatment simulations,
superimposed on each other.
Both sets of data come from simulations performed at the density $ \rho = 0.1 \, \ANG^{-3} $.
The two curves are nearly identical in form.
There are very small deviations between the two,
and these are the most pronounced around the first shell.
Define $ \Delta g(r) $ to be
the radial pair distribution function of the perturbative treatment simulation
subtracted from
that of the full inclusion treatment simulation.
We plot $ \Delta g(r) $ around the first shell in the inset of Fig.~\ref{fig:paper2_fig10}.
The inclusion of the three-body interaction during sampling
softens the pair distribution function,
and we can see this especially in the first shell,
which becomes noticeably flatter and wider.
Note that we show this example for the highest density explored in this work,
$ \rho = 0.1 \, \ANG^{-3} $,
because it is the case for which the change in the structure is the most pronounced.
Similar behaviours occur at lower densities,
but with a smaller change in the distribution.

\begin{figure} [H]
    \centering
    \includegraphics[width=\linewidth]{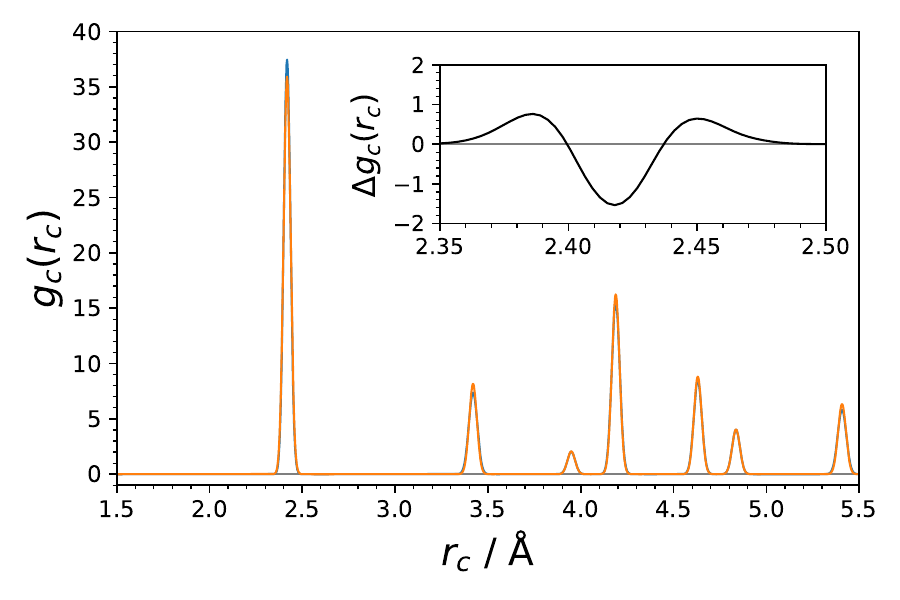}
    \caption{
        The radial centroid pair distribution function $g_c(r_c)$
        of solid parahydrogen at a density $\rho = 0.1 \, \ANG^{-3}$,
        given by Eq.~(\protect\ref{eq:results:pairdist:centroid_pairdist}).
        The simulation labels are identical to those in the caption of
        Fig.~\protect\ref{fig:paper2_fig10}.
        In the inset we show $\Delta g_c(r_c)$,
        which if the difference between the orange and blue curves
        in the region around the first shell of the two $g_c(r_c)$ distributions.
    }
    \label{fig:paper2_fig11}
\end{figure}

In Fig.~\ref{fig:paper2_fig11},
we show the corresponding $ g_c(r_c) $ curves
under the same conditions and context of Fig.~\ref{fig:paper2_fig10}.
The centroid radial pair distribution is much more localized than its
corresponding radial pair distribution function.
As was the case for the $ g(r) $ functions,
the perturbative and full inclusion treatment simulations produce very similar $ g_c(r_c) $ functions.
The differences are most easily seen around the first peak.
We define $ \Delta g_c(r_c) $ similarly to how we defined $ g(r) $ earlier,
and plot it around the first peak in the inset of Fig.~\ref{fig:paper2_fig11}.
Once again,
we see that the three-body interaction softens the peak
in the first shell of the distribution.

We can use our findings thus far
to describe the change in the lattice structure
when the three-body interaction is introduced.
Recall from Sec.~\ref{sec:results:centroid} that
the three-body interaction has a negligible effect on the centroid bead distribution,
even up to $ \rho = 0.1 \, \ANG^{-3} $.
Also, we see from Figs.~\ref{fig:paper2_fig10} and
\ref{fig:paper2_fig11}
that both the radial pair distribution and the centroid radial pair distribution
soften around the first shell after including the three-body interaction in the sampling.
These findings indicate that,
at least up to the densities explored in this paper,
the three-body interaction
does not increase the overall quantum mechanical ``spreading'' of each individual particle,
but it does give each particle more freedom to move around its nominal lattice site.

We should admit, of course,
that the arguments presented here
about the effect of the three-body interaction on the solid's structure
come from a model that fails to produce the experimental EOS.
However,
these structural changes are consequences of softening the overall interaction energies
of the solid at higher densities.
Thus,
even if we had used a model that produced the correct EOS
(\textit{i.e.} one that shifts the FSH-[2B] pressure-density curve down to the experimental data instead of shooting past it)
we would likely have seen similar changes in the structure, qualitatively.
The radial and centroid radial pair distributions would both still have softened,
but the changes would be less pronounced.

\begin{figure} [H]
    \centering
    \includegraphics[width=\linewidth]{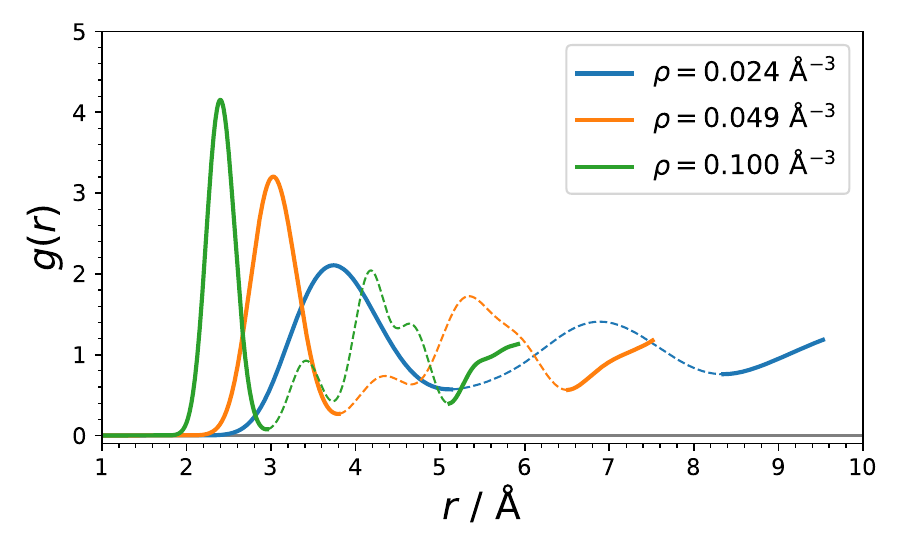}
    \caption{
        The radial pair distribution function $g(r)$ of solid parahydrogen,
        collected from FSH-[2B] simulations
        at densities of
        $\rho = 0.024 \, \ANG^{-3}$ (blue curve),
        $\rho = 0.049 \, \ANG^{-3}$ (orange curve), and
        $\rho = 0.1   \, \ANG^{-3}$ (green curve).
        The dashed portion of each curve is a visual guide for the reader.
        It highlights the second shell region of the blue curve,
        as well as the corresponding ``second shell'' regions in the orange and green curves,
        when the change in density is accounted for.
    }
    \label{fig:paper2_fig12}
\end{figure}

We can explore how the structure of the lattice changes as a function of density.
In Fig.~\ref{fig:paper2_fig12},
we show the pair distributions $ g(r) $
for three different densities,
collected from perturbative treatment simulations.
A change in the density has a far greater effect on the radial pair distribution
than does the presence of the three-body interaction potential.
As the density increases,
the first shell becomes narrower and more peaked.
Features of the pair distribution function that are hidden at low densities
by the quantum delocalization of the parahydrogen molecules
become visible at higher densities.
For example,
focus on the regions indicated by the dashed lines in Fig.~\ref{fig:paper2_fig12}.
At a density of $ \rho = 0.024 \, \ANG^{-3} $,
this region appears as a single, uniform shell.
As the density increases,
the shell develops sharper, more distinguishable features.

\section{Conclusions} \label{sec:conclusion}
We used PIMC simulations
to study the EOS of solid parahydrogen at $ T = 4.2 $ K
using \textit{ab initio} pair and three-body interaction potentials.
Periodic boundary conditions were accounted for during the simulations.
The finite system size errors were corrected using potential tail corrections,
and the Trotter factorization errors were accounted for
either by extrapolation or by using a sufficiently small imaginary time step.

We found that the difference between
the perturbative and full inclusion treatments of the three-body interaction
is fairly small, even at the highest densities explored in this paper.
The total energies and pressures in both cases are very similar.
The full inclusion of the three-body interaction
in the PIMC sampling protocol
causes noticeable changes in each of the
kinetic, pair potential, and three-body potential energies.
However, when added together,
these individual changes largely cancel out,
resulting in only a small decrease in the total energy.
This behaviour can be explained by the fact that
the inclusion of the three-body interaction
has only minor effects on the structure of the solid.
For example,
the centroid bead distribution shows negligible change
going from the perturbative to the full inclusion case.
The radial and centroid radial pair distributions soften very slightly,
with the majority of the change in both cases occurring around the first peak.
The success of the perturbative method
shows that the three-body interaction,
and possibly also higher-order many-body interactions,
can often be treated as a sort of density-dependent background energy.

In Sec.~\ref{sec:results:centroid},
we show how the centroid bead distribution changes as a function of density.
It will be of interest to test the deconvolution procedure derived to connect centroid radial distributions to real space ones in the context of liquid parahydrogen,\cite{blinov2004connection} and  see how these concepts can be applied to a quantum solid.
The centroid bead distribution
is a measure of the quantum delocalization of the molecules,
and thus it could be used to help predict \textit{a priori}
the coupling matrix for
numerical coarse-grained path-integral simulations of solid parahydrogen.\cite{roy2002centroid,convolve:19ryu}

The pair potential on its own is too repulsive at high densities
to successfully reproduce the EOS.
The three-body interaction,
when used as the only many-body correction energy term,
overcorrects the effects of the repulsive wall of the pair interaction.
The resulting pressure-density curve severely underestimates experimentally observed pressures.
It is possible that,
to perform accurate simulations,
we need four-body and higher-order many-body interaction terms.
It could also be the case that
static interaction potentials on their own are inadequate
for simulating solid parahydrogen at high densities. A final potential source of inaccuracy is the use of the AHR approximation where parahydrogen molecules are treated as point-like particles. One way to remedy this situation would be to explicitly include molecular rotations in our path integral description. Such an extension is possible via the use of established rotational path integral techniques.\cite{cui1997rotational,marx1999path,pathinteg:04blin,pathinteg:05blin,superfluidh2:13zeng:1,superfluidh2:13zeng:2,zeng2014microscopic,zeng2016moribs}

\section*{Acknowledgements}
The authors acknowledge the Natural Sciences and Engineering Research Council (NSERC) of Canada (RGPIN-2016-04403), the Ontario Ministry of Research and Innovation (MRI), the Canada Research Chair program (950-231024), and the Canada Foundation for Innovation (CFI) (project No. 35232). A.~I. acknowledges the support of the NSERC of Canada (CGSD3-558762-2021).

\section*{DATA AVAILABILITY}
The data that support the findings of this study are available from the corresponding author upon reasonable request.

\section*{References}

\bibliography{./biblio}
\bibliographystyle{ieeetr}

\end{document}